# Digital synthesis of histological stains using micro-structured and multiplexed virtual staining of label-free tissue

Short Title: Multiplexed virtual staining of label-free tissue

**Yijie Zhang [†,1,2,3], Kevin de Haan [†,1,2,3], Yair Rivenson[†,1,2,3,*], Jingxi Li[1,2,3], Apostolos Delis[4], Aydogan Ozcan[1,2,3,5,*]**

[1]Electrical and Computer Engineering Department, University of California, Los Angeles, CA, 90095, USA.

[2]Bioengineering Department, University of California, Los Angeles, CA, 90095, USA.

[3]California NanoSystems Institute (CNSI), University of California, Los Angeles, CA, 90095, USA.

[4]Department of Computer Science, University of California, Los Angeles, CA, 90095, USA.

[5]Department of Surgery, David Geffen School of Medicine, University of California, Los Angeles, CA, 90095, USA.

† Equally contributing authors

* Corresponding authors: rivensonyair@ucla.edu ; ozcan@ucla.edu

**Abstract**

Histological staining is a vital step used to diagnose various diseases and has been used for more than a century to provide contrast to tissue sections, rendering the tissue constituents visible for microscopic analysis by medical experts. However, this process is time-consuming, labor-intensive, expensive and destructive to the specimen. Recently, the ability to virtually-stain unlabeled tissue sections, entirely avoiding the histochemical staining step, has been demonstrated using tissue-stain specific deep neural networks. Here, we present a new deep learning-based framework which generates virtually-stained images using label-free tissue, where different stains are merged following a micro-structure map defined by the user. This approach uses a single deep neural network that receives two different sources of information at its input: (1) autofluorescence images of the label-free tissue sample, and (2) a "digital staining matrix" which represents the desired microscopic map of different stains to be virtually generated at the same tissue section. This digital staining matrix is also used to virtually blend existing stains, digitally synthesizing new histological stains. We trained and blindly tested this virtual-staining network using unlabeled kidney tissue sections to generate micro-structured combinations of Hematoxylin and Eosin (H&E), Jones silver stain, and Masson's Trichrome stain. Using a single network, this approach multiplexes virtual staining of label-free tissue with multiple types of stains and paves the way for synthesizing new digital histological stains that can be created on the same tissue cross-section, which is currently not feasible with standard histochemical staining methods.

## Introduction

Histological analysis is used to diagnose a wide variety of diseases. It is considered the gold standard for tissue-based diagnostics, with some of the well-established versions of common stains such as Hematoxylin and Eosin (H&E) having been used for over a hundred years[1]. The histological staining process first requires slicing of the fixed tissue specimen into 2-10 micron sections, which are fixed to microscope slides. Histological staining chemically introduces contrast to tissue sections, which can then be analyzed and used to screen for diseases through brightfield microscopic imaging of the stained samples. However, histological staining can be a long and labor-intensive process, particularly in the case of special stains such as e.g., Jones silver stain and Masson's Trichrome stain. Therefore, the tissue staining process can increase both the time to diagnosis as well as the associated costs.

A wide variety of stains have been developed over the years in order to enable the visualization of different targeted tissue constituents. For example, Hematoxylin stains cell nuclei, while Masson's trichrome stain is used to view connective tissue[2]. These stains have also been chemically mixed to enable the visualization of different biomarkers. An example of this is when the Periodic acid-Schiff (PAS) and Alcian-blue stains are used in conjunction to perform differential staining of glycoproteins[3].

In recent years, various methods have been developed to replace the requirement for histochemical staining of the sample, in an attempt to avoid (1) the destructive nature of the labeling process on the specimen, allowing tissue preservation for more advanced analysis; (2) the lengthy and laborious labelling steps, saving time and cost; and (3) unnecessary additional biopsies from the same patient due to tissue depletion. Some of the earlier alternative contrast generation methods utilize various processes that result from light matter interaction including e.g., nonlinear microscopy[4], Raman scattering[5], programmable supercontinuum pulses[6] and reflectance confocal microscopy[7]. As pathologists (and more recently, machine learning algorithms) are trained to mainly perform diagnoses using histologically stained specimen, images that were generated by alternative contrast mechanisms might require additional training. Recent efforts have also focused on the development of computational methods to create brightfield microscopy images that closely resemble the stained versions of the same specimens. For example, digitally generated pseudo-stains were demonstrated using analytical and statistical learning-based approaches that transform an input pixel (or pixel spectrum) into an RGB output pixel[5,8,9]. Some of these pixel-to-pixel transformation approaches have also used rapid staining methods to provide contrast to cell nuclei[4,9].

Recently emerging deep learning methods now allow algorithms to learn accurate transformations between many different imaging modalities[10–14]. Notably, by utilizing the statistical correlations between the structures in images of unstained tissue slides to the structures in images of the same slides once stained, an unstained tissue sample can be virtually stained by a trained deep neural network, without the need for any chemical processing. For example, using deep learning, autofluorescence images of unlabeled tissue samples have been virtually stained with different types of stains[15]. These virtual stains were validated through a blind study by a team of board-certified pathologists to reveal that there is no statistically significant difference in quality between a virtually stained image and a standard histochemically stained version of the same sample imaged with a brightfield microscope, in terms of both the stain quality and the diagnostic information. Various other techniques to perform virtual staining of unlabeled tissue have been demonstrated by e.g., using quantitative phase images[16], or a combination of two photon excitation and fluorescence lifetime imaging[17]. Researchers have also used deep learning to improve the accuracy of diagnosis using H&E images[18]; it has been shown that deep neural networks can be used to normalize stains, making them more consistent, which allows the automated diagnostic analysis to be performed more easily[19].

In this paper, we demonstrate a novel machine learning-based framework which allows users to virtually create micro-structured and multiplexed histological stains on the same tissue section using only a *single* artificial deep neural network. Using this technique, a trained deep neural network can (1) perform virtual staining upon a defined region of interest following a micro-structure map that is user defined, and (2) perform blending of multiple virtual stains and synthesizing new digital stains. This framework uses the stain type as the input class for a conditional generative adversarial network (GAN) to transform input images, consisting of two autofluorescence images of an unlabeled tissue sample, into a virtually-stained image of the same label-free sample. To do this, we introduce a "digital staining matrix" which is used as part of the input to the deep network, which encodes in space the stain type, i.e., each pixel can be virtually stained using a different stain type or a different set of histological stains (see Figure 1).

To demonstrate the utility of this technique, we trained a single neural network to virtually stain autofluorescence images of unlabeled kidney needle core biopsy tissue sections with H&E, Jones silver stain, and Masson's Trichrome stain, following a user defined micro-structure map as illustrated in Figure 1. Synthesizing different histological stains and their combinations, following a user defined micro-structure at the same tissue section is currently not feasible with standard histological staining process, where different stains are histochemically processed at *different* tissue sections, depleting tissue. Our approach entirely eliminates the need for this, preserving tissue for further analysis, while also paving the way for synthesizing new digital histological stains on the same tissue section, on demand.

## Results

As summarized in Figure 1, we demonstrate a method which can be used to perform virtual staining of unlabeled tissue sections using two channels of tissue autofluorescence along with a digital staining matrix, used as inputs to a trained deep neural network. We chose to demonstrate the framework using kidney tissue and three different stains: H&E, Masson's Trichrome, and Jones stain, as they are jointly used for practical renal disease diagnostics. Visualizations of the comparisons between the histochemically and virtually stained tissue sections can be seen in Figures 1 and 2.

We further validated the accuracy of the network inference using the structural similarity index[20] (SSIM), which is defined as:

$$SSIM(a,b) = \frac{(2\mu_a\mu_b + C_1)(2\sigma_{a,b} + C_2)}{({\mu_a}^2 + {\mu_b}^2 + C_1)({\sigma_a}^2 + {\sigma_b}^2 + C_2)} \qquad (1)$$

where $\mu_a$ and $\mu_b$ are the averages of *a* and *b*, the two images being compared, $\sigma_a$ and $\sigma_b$ are the standard deviations of *a* and *b*, and $\sigma_{ab}$ is the cross covariance of *a* and *b*. $C_1$ and $C_2$ are stabilization constants which are used to avoid division by zero.

Table 1 reports the average SSIM across four unique blindly-tested kidney tissue blocks, each coming from a different patient. Each one of these blocks was in turn made up of 16 to 60 patches (1224×1224 pixels, or 0.16 $mm^2$ per patch), each made up of an unlabeled autofluorescence image pair and its co-registered histochemically stained counterpart (see the Methods section). As this comparison relies on histochemical staining of the same tissue section, a different section from each tissue block was used for each of the three different stain types. The variation in the number of patches is partially due to the tissue blocks varying in size among patients; furthermore, the images that could not be successfully co-registered due to e.g., histochemical staining induced tissue distortions were excluded from the SSIM calculations. Three different SSIM values are calculated for each stain type in order to prove that this new virtual staining technique is successful: (1) The SSIM between the conditional multistain network output image and the corresponding histochemically stained tissue image; (2) The SSIM between the output of a previously validated[15] single stain network architecture (see Methods section) and the corresponding histochemically stained tissue; (3) The SSIM between the outputs of the multistain network and the single stain network for each of the three stains. As shown in Table 1, a high structural similarity is found for all three cases. Furthermore, the SSIM values calculated for cases (1) and (2) are found to be very similar, which indicates that the images generated by the multistain network achieve the same virtual staining performance as was previously reported and validated using a single stain network[15]. The particularly high structural similarity between the two different virtual staining techniques (SSIM calculation 3) is further important since they have perfect co-registration with respect to each other as they use the same raw fluorescence images. Together, these results suggest that the multistain network generates highly accurate virtually stained images.

These different set of comparisons between SSIM values are required, as the SSIM values between any virtually stained and histochemically stained images are dependent on a number of factors, some of which are external to the performance of the trained neural network. *Perfect* co-registration is not feasible, particularly since physical changes are made to the tissue sample during the actual staining process,[15] partially lowering the structural similarity values regardless of the success of the virtual staining network. Furthermore, one of the major benefits of the deep learning-based virtual staining approach is stain normalization, as the network output will not exhibit the staining variability of standard histochemical staining process performed e.g., by histotechnologists[15]. While this is certainly a desired feature and will help to improve the consistency of diagnoses, it lowers the SSIM values due to histotechnologist-to-histotechnologist variations that are encountered in our ground truth images.

As another quantitative metric, next we compared the average percentage differences of brightness and chroma components (using the YCbCr color space) for the three cases reported in Table 1, i.e., (1) Multistain network output vs. histochemically stained tissue; (2) Single stain network output vs. histochemically stained tissue; and (3) Multistain network output vs. single stain network output. As summarized in Table 2, similar to the case of SSIM, the color differences for cases (1) and (2) are very similar. Case (3) has particularly small differences, indicating that the two networks (multistain vs. single stain) behave very similarly. The brightness (Y) of the multistain network has a relatively low change with respect to the image of the histochemically stained tissue. This ranges from 3.84% to 8.57% depending on the stain. The color distances (Cb and Cr differences) are even smaller (ranging from 0.51% to 2.60% depending on stain type, see Table 2), indicating that the multistain network accurately generates the correct colors that represent each stain. Together, these results further demonstrate that the multistain network is capable of accurate virtually staining of unlabeled autofluoresence images of tissue samples, and that the output of the multistain network matches the accuracy of a previously validated tissue and stain specific neural network[15].

One of the major advantages to using a class conditional neural network is that it can perform micro-structured virtual staining of tissue sections. Using the presented method, virtual staining of specific areas or structures within the tissue sections is performed by staining different areas of the tissue according to a given micro-structure map. The digital staining matrix, which defines the microstructure map to virtually perform different stains for each of these areas can be generated manually, or through the use of a computer algorithm that selects structures based on some diagnostic criteria. An example of virtual stain micro-structuring according to a manually drawn micro-structure map is shown in Figure 3. In this example, marked areas are virtually stained with Masson's trichrome and Jones stains, while the remaining areas not selected are stained with H&E. A co-registered image of the same field of view (FOV) imaged after histochemical H&E staining is also shown for comparison.

Stain blending can also be used to digitally synthesize new types of stains. Rather than using the digital staining matrix to generate individual stains, a mixture of multiple stains can be chosen. This stain mixture is generated by having the digital staining matrix mix two or more stains in the desired tissue areas, simultaneously and at controllable ratios (see Figure 4). In other words, the newly generated stain can be tuned on demand by simply changing the ratio between the different values in the digital staining matrix, making the different stain combinations more or less pronounced. Figure 4 demonstrates several of these stain combinations for different pairs of stains. By using these blended stains, aspects of the different stains become visible at the same time, which may allow pathologists to more easily view different tissue structures and perform diagnosis. For example, Figure 4 (a-e) demonstrate blending between H&E and Jones silver stain; H&E allows for easy differentiation of cell nuclei and cytoplasms[21], while Jones silver stain gives contrast to basement membranes[22]. By blending the two stains, aspects of these two stains are shown simultaneously.

While the digital stains generated here are completely unique to virtual staining, various histochemical stains have also been mixed together to generate new stain combinations[3]. However, developing these new stain combinations chemically can take a large amount of time and resources to mature. In contrast, the stain blending combinations presented here can be developed on demand by simply changing the values of the digital staining matrix until the desired stain is achieved. The virtual stain blending presented here also has a different effect upon the tissue than standard histochemical stains, allowing for a new mode of micro-structured visualization for pathologists.

## Discussion

In this paper we demonstrated that autofluorescence images of a label-free tissue sample can be used to perform micro-structured and multiplexed virtual staining using a deep neural network. By adding a digital staining matrix to the input of the neural network, we can generate multiple virtual stains on the same tissue section using a *single* network. The success of this approach has been validated using kidney tissue sections and three different stains – H&E, Masson's Trichrome and Jones stain – allowing a pathologist to view the same areas of the sample with all three stains, perfectly matched at the same tissue cross-section. The digital staining matrix also allows us to perform micro-structured virtual staining of a label-free sample, where the sub-area for each stain can be defined either manually or using a separate algorithm. This approach can further be used to perform stain blending by using a digital combination of the stains that the multistain neural network is trained for.

The ability to perform multiple stains on a single tissue section using a single neural network alongside the newly added capabilities of stain blending, synthesis, and micro-structured virtual staining have the potential to improve the accuracy and consistency of tissue-based diagnoses. These new techniques might allow the pathologist to get more relevant information from the tissue than is otherwise possible. By applying stains to specific areas, each tissue constituent can be stained with the most relevant stain. By blending stains, the network is able to simultaneously display information contained in each of the separate stains, giving additional channels of information to the pathologists making diagnoses.

These virtual staining techniques also open up opportunities to augment the diagnostic workflow currently used by pathologists and/or machine learning-based diagnostic algorithms. Virtual staining normalizes the stain quality, improving its consistency and removing the variations (caused by e.g., the manual histochemical staining performed by trained professionals) not learned by the neural network[15]. Furthermore, micro-structured staining and stain blending can ensure that the diagnosis platform has the most relevant information possible, reducing the amount of unnecessary data viewed/processed by either a pathologist or an algorithm. In this regard, we believe that the push-pull relationship between the presented virtual staining framework and diagnosticians (human or AI-based) would lead to new uses of the capabilities of this unique framework in pathology and clinical diagnosis, which must all be clinically validated through rigorous testing and blinded large-scale studies.

## Methods:

### Data acquisition

Unstained formalin-fixed and paraffin-embedded (FFPE) kidney tissues were sectioned into thin, 2µm slices and fixed on a standard microscope glass slide. These tissue sections were obtained under UCLA IRB 18-001029. Using a conventional widefield fluorescence microscope (IX83, Olympus) equipped with a 20×/0.75 NA objective lens (Olympus UPLSAPO) and with two separate filter cubes, DAPI (OSFI3-DAPI-5060C, EX 377/50 nm EM 447/60 nm, Semrock) and Texas Red (OSFI3-TXRED-4040C, EX 562/40 nm EM 624/40 nm, Semrock), these unlabeled tissue sections' autofluorescence was imaged. The tissue sections were neither deparaffinized or cover-slipped before being imaged by the fluorescence microscope. The exposure time for the DAPI channel was 50 ms and for the Texas Red channel was 300 ms. Once the autofluorescence of the specimen was imaged, the slides were histochemically stained using standard H&E or Jones or Masson's trichrome stains and were then cover-slipped. The staining of the slides was performed by the UCLA Translational Pathology Core Laboratory (TPCL). These histochemically stained slides were then imaged using a scanning microscope (Aperio AT, Leica Biosystems, 20×/0.75NA objective with a 2× adapter) to create the target labels used to train, validate and test our neural network models.

We used the two autofluorescence images of the unlabeled tissue in conjunction with a digital staining matrix which selects the stain or set of stains to generate as the input to a neural network. This input is transformed using a class conditional generative adversarial network (c-GAN) into an equivalent image of a stained tissue section of the same field-of-view.

As the deep neural network aims to learn the transformation from autofluorescence images of the unlabeled tissue specimen to those of a stained specimen, it is crucial that the FOVs are accurately aligned. Furthermore, when more than one autofluorescence channel is used as the network's input, the various filter channels must be aligned. In order to use three different stains (H&E, Masson trichrome and Jones), we implemented the image pre-processing and alignment for each input and target image pair from those three staining datasets respectively.

The registration steps matching the autofluorescence and brightfield images follow the process reported by Rivenson *et al*.[15] It begins by performing a rough global registration between the fluorescence and brightfield images, and progressively aligns them at smaller scales until subpixel-level co-registration is achieved. One major addition is that when using multiple autofluorescence channels as the network input (i.e. DAPI and Texas Red), they must be aligned even though the images from the two channels are captured using the same microscope; the corresponding FOVs from the two channels are not precisely aligned to the subpixel-level, particularly on the edges of the FOVs. Therefore, we have applied an elastic pyramidal registration algorithm to accurately align the multiple autofluorescence channels. This elastic registration algorithm matches the local features of the two image channels by hierarchically breaking the images into smaller and smaller blocks, and then matching the corresponding blocks[10]. The calculated transformation map was then applied to the Texas Red images to ensure that they are aligned to the corresponding images from the DAPI channel. Finally, we stitch the aligned images from two channels and get aligned whole slide images of the sample which contain both the DAPI and Texas Red channels.

Before feeding the aligned pairs into neural network, we implement normalization on the whole slide images of the DAPI and Texas Red images respectively. This whole slide normalization is performed by subtracting the mean value of the entire tissue sample and dividing it by the standard deviation among pixel values, with the background regions excluded from the calculations of the mean and standard deviation.

**Deep neural network architecture, training and validation**

In this study, we used a class conditional GAN architecture to learn the transformation from label-free unstained autofluorescence input images to the corresponding bright-field image using three different stains (H&E, Masson trichrome and Jones). Following the co-registration of the autofluorescence images to the bright-field images, these accurately aligned FOVs were randomly partitioned into overlapping patches of 256×256 pixels, and further augmented through rotation and flipping. The patches were then used to train the GAN. During the training process, this class conditional GAN uses a set of one-hot encoded matrices, referred to as a digital staining matrix, which is concatenated to the network's 256×256 input image / image stack patches, where each matrix corresponds to a different stain. One way to represent this conditioning is given by:

$$\tilde{c} = [c_1, \ c_2, \ c_3] \qquad (2)$$

where $[\cdot]$ refers to concatenation, and $c_i$ represents a 256×256 matrix for the label for the $i$-th staining type (in this example: H&E, Masson trichrome and Jones). For an input and target image pair from the $i$-th staining dataset, the $c_i$ is set to be an all ones matrix, while all other remaining matrices are assigned zero values accordingly.

The GAN is composed of two deep neural networks, a generator and a discriminator (Figure 5). During the GAN training, the generator learns to perform a statistical transformation and generate the virtually stained image, while the discriminator attempts to distinguish between the histochemically stained images and their virtually stained counterparts. The networks improve by learning from one another, improving the quality of the virtually stained images. For this task, we define the loss functions of the generator and discriminator to be:

$$\ell_{generator} = L_1\{z_{label}, G(x_{input}, \tilde{c})\} + \lambda \times TV\{G(x_{input}, \tilde{c})\} + \alpha \times \left(1 - D\left(G(x_{input}, \tilde{c}), \tilde{c}\right)\right)^2 \quad (3)$$

$$\ell_{discriminator} = D\left(G(x_{input}, \tilde{c}), \tilde{c}\right)^2 + \left(1 - D(z_{label}, \tilde{c})\right)^2$$

where the total variation ($TV$) operator and mean absolute error ($L_1$ norm) are used to regularize the generator's output and ensure that it is highly accurate. They are defined as:

$$TV(z) = \sum_p \sum_q |z_{p+1,q} - z_{p,q}| + |z_{p,q+1} - z_{p,q}| \quad (4)$$

$$L_1(z, \ G) = \frac{1}{P \times Q} \sum_p \sum_q |z_{p,q} - G(x_{input}, \tilde{c})_{p,q}| \quad (5)$$

where $D(\cdot)$ and $G(\cdot)$ refer to the discriminator and generator network outputs, respectively, $z_{label}$ denotes the bright-field image of the histochemically stained tissue, and $x_{input}$ represent the input to the neural network. P and Q represent the number of vertical and horizontal pixels for the image patch, and p and q represent the pixel locations. The regularization parameters ($\lambda$, $\alpha$) were set to be 0.02 and 2000, respectively, which accommodate the total variation loss term to be approximately 2% of the $L_1$ loss and the discriminator loss term to be 98% of the total generator loss.

The generator follows a modified version of the U-net architecture[23] which is visualized in Figure 5 (a). This U-net consists of four "down-blocks" followed by four "up-blocks". Each of the down-blocks is made up of three convolutional layers and their activations functions, which together double the number of channels. These convolutional layers are followed by an average pooling layer with a stride and kernel size of two, which effectively down-samples the image. The up-blocks first bilinearly resize the tensors, up-sampling them by a factor of two. This is then followed by three convolutional layers and their activation functions. These convolutional layers together reduce the number of channels by a factor of four. Between each of the up and down blocks of the same level, a skip connection is used. These skip connections concatenate the output of the down-blocks with the up-sampled values, allowing data to be passed at each level. Following these down- and up-blocks, a convolutional layer is used to reduce the number of channels to three, which correspond to the three color channels in the brightfield image.

The discriminator network visualized in Figure 5 (b), receives six input channels. Three channels (YCbCr color map) come from either the generator output or the target/label and three from the one-hot encoded digital staining matrix. The discriminator architecture, consists of a convolution layer which transforms this input into a 64-channel feature map and in turn passes through a set of five blocks, each consisting of two convolution layers and their corresponding activation functions. The second of these convolutional layers doubles the number of channels and has a stride of two. These five blocks are followed by two fully connected layers, which reduce the dimensionality to a single number, which is acted upon by a sigmoid activation function.

The convolution filter size throughout the GAN was set to be $3 \times 3$, which are acted upon by the leaky-ReLU activation function, described as:

$$LeakyReLU(x) = \begin{cases} x & for\ x > 0 \\ 0.1x & otherwise \end{cases} \quad (6)$$

The learnable parameters were updated through the training using adaptive moment estimation (Adam) optimizer with a learning rate of $1 \times 10^{-4}$ for the generator network and $2 \times 10^{-6}$ for discriminator network. For each discriminator training step, there were ten iterations of the generator network. The batch size of the training was set to be 8.

**Virtual staining of unlabeled tissue with a single stain**

Once the network is trained, the one-hot encoded label $\tilde{c}$ is used to condition the network to generate the desired stained images. In other words, a matrix $c_i$ is set to be an all ones matrix and other remaining matrices are set to be all zeros to solely generate the $i$-th stain.

**Stain blending and micro-structured virtual staining of unlabeled tissue**

Following the training process of the neural network model, we can use the conditional matrices in manners different than to those it was trained for in order to virtually create new types of stains. The basic encoding rule that should be satisfied can be summarized as:

$$\sum_{i=1}^{N_{stains}} c_{i,j,k} = 1 \qquad (7)$$

In other words, for a given set of indices, *j,k*, the sum across the number of stains that the network was trained with ($N_{stains} = 3$ in our example) should be equal to 1. By modifying the class encoding matrices to use a mixture of multiple classes, the various stains can be blended, creating unique stains with features emanating from the various stains learned by the artificial neural network. Examples of these blended stains are illustrated in Figure 4.

Another possible use of our trained multiplexed staining neural network is to partition the tissue field-of-view into different regions of interest (ROI-s), with each one of them virtually stained using different specific stains, or the blend of a sub-set of these stains:

$$\sum_{i=1}^{N_{stains}} c_{i,j,k} = 1 \quad for \quad j,k \subseteq ROI \qquad (8)$$

where *ROI* is the defined region-of-interest in the sample field of view. Multiple non-overlapping *ROI*-s can be defined across a field-of-view, with different stains applied to different regions of interest or micro-structures. These can either be user defined, or algorithmically generated. As an example, a user can manually define various tissue areas via a graphical user interface and stain them with different stains. This results in different tissue constituents being stained differently, as illustrated in Figures 1 and 3. We have implemented this ROI selective staining (micro-structured staining) using the Python segmentation package Labelme[24]. Using this package, we generate logical masks according to labeled ROIs, which are then processed to be the $\tilde{c}_{ROI}$ label for specific microscopic areas. Other manual, software or hybrid approaches can be used to implement selection of other tissue structures.

**Single stain network used for SSIM calculations**

To generate virtually stained images using the single stain network, a network with the same architecture excluding the digital staining matrix was used. A separate network was trained for each of the three stains, using the portion of the dataset specific to that stain. This single stain network followed the approach previously reported[15].

**Implementation details**

The virtual staining network was implemented using Python version 3.6.0, with TensorFlow framework version 1.11.0. We implemented the software on a desktop computer with an Intel Xeon W-2195 CPU at 2.30 GHz and 256GB RAM, running a Microsoft Windows 10 operating system. Network training and testing were performed using a single NVDIA GeForce RTX 2080 Ti GPU. The network was trained for 21000 discriminator training steps over 47 hours. Using a single GPU, inference can be performed at a rate of 3.9 s per 1 $mm^2$ of unlabeled tissue.

**Acknowledgements**

The authors acknowledge the funding of NSF Biophotonics Program.

## Tables and Figures

***Table 1:*** *Comparison of SSIM values among the different networks and the histochemically stained tissue images. The averages and standard deviations are calculated across the four measured tissue sections.*

| Stain type | (1) Multistain network output vs. histochemically stained tissue | | (2) Single stain network output vs. histochemically stained tissue | | (3) Multistain network output vs. single stain network output | | Total number of image patches compared |
|---|---|---|---|---|---|---|---|
| | Average | Standard Deviation | Average | Standard Deviation | Average | Standard Deviation | |
| H&E | 0.898 | 0.021 | 0.905 | 0.022 | 0.967 | 0.006 | 198 |
| Masson's Trichrome | 0.850 | 0.011 | 0.855 | 0.023 | 0.942 | 0.010 | 207 |
| Jones stain | 0.803 | 0.007 | 0.803 | 0.010 | 0.917 | 0.007 | 118 |

***Table 2:*** *Comparison of brightness and chroma differences (using the YCbCr color space) between (1) Multistain network output and histochemically stained tissue; (2) Single stain network output and histochemically stained tissue; and (3) Multistain network output and single stain network output. The averages and standard deviations are calculated across the four measured tissue sections.*

| Stain type | Comparison | Y difference (%) | | Cb difference (%) | | Cr difference (%) | | Total number of image patches compared |
|---|---|---|---|---|---|---|---|---|
| | | Average | Standard Deviation | Average | Standard Deviation | Average | Standard Deviation | |
| H&E | (1) Multistain network vs. histochemically stained tissue | 6.62 | 3.32 | 0.51 | 0.18 | 1.69 | 1.18 | 198 |
| | (2) Single stain network vs. histochemically stained tissue | 7.78 | 3.48 | 0.87 | 0.21 | 2.04 | 1.41 | |
| | (3) Multistain network vs. single stain network | 1.48 | 0.12 | 0.22 | 0.03 | 0.72 | 0.20 | |
| Masson's Trichrome | (1) Multistain network vs. histochemically stained tissue | 3.85 | 1.50 | 1.34 | 0.87 | 2.60 | 1.16 | 207 |
| | (2) Single stain network vs. histochemically stained tissue | 5.31 | 1.32 | 2.09 | 1.51 | 3.00 | 1.44 | |
| | (3) Multistain network vs. single stain network | 1.96 | 1.70 | 0.43 | 0.19 | 1.35 | 0.53 | |
| Jones stain | (1) Multistain network vs. histochemically stained tissue | 8.56 | 2.01 | 0.82 | 0.12 | 2.45 | 0.69 | 118 |
| | (2) Single stain network vs. histochemically stained tissue | 9.07 | 1.93 | 1.33 | 0.21 | 3.15 | 0.86 | |
| | (3) Multistain network vs. single stain network | 4.32 | 1.01 | 0.34 | 0.11 | 1.15 | 0.24 | |

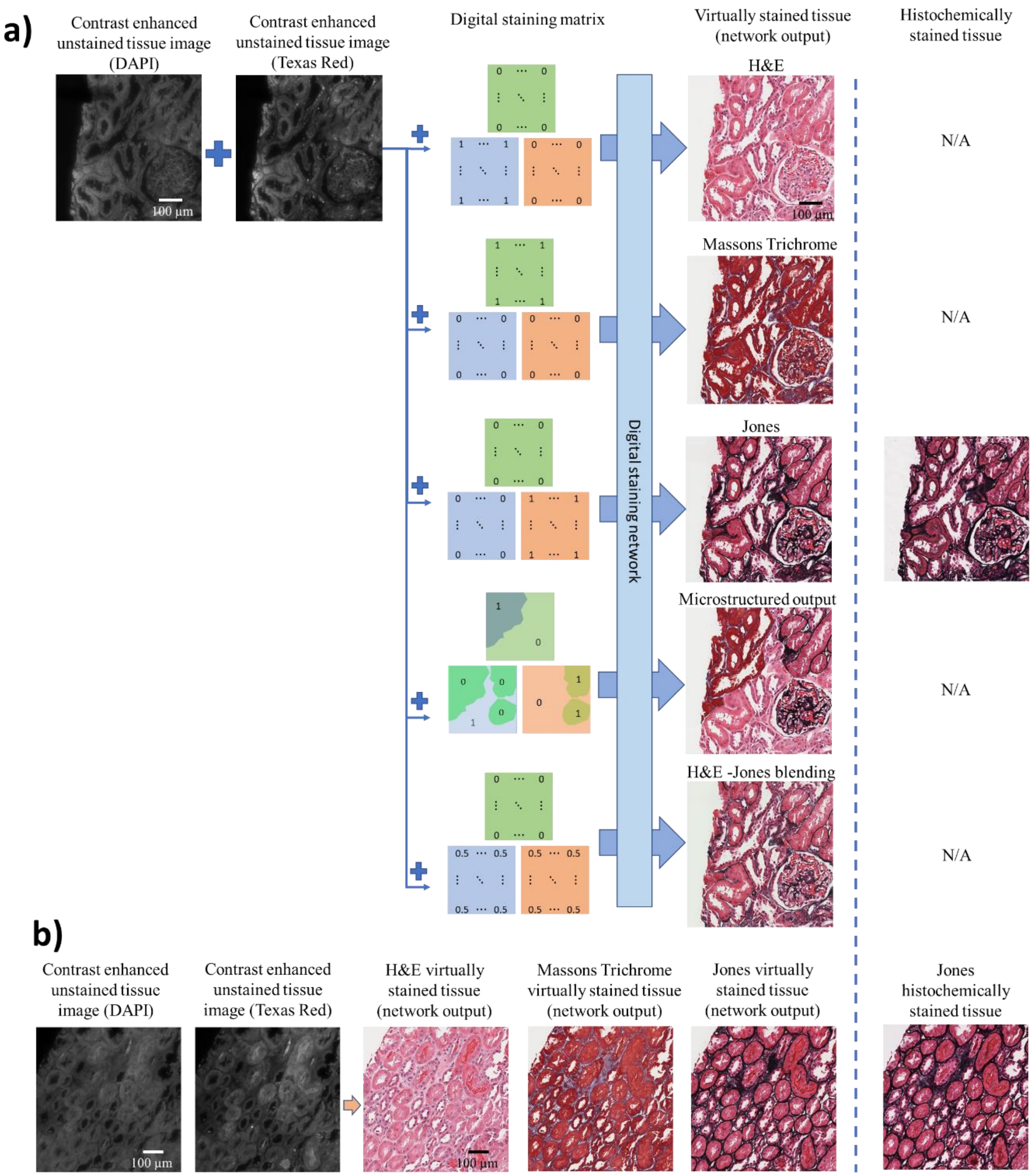

***Figure 1:*** *Demonstration multiple stains being virtually generated using a class conditional neural network and the two autofluorescence channels (DAPI and Texas Red) of the label-free tissue sample. (a) Steps involved in virtually creating the various stains. By adding a class condition to the network using a digital staining matrix, a single network can be used to generate multiple stains, or a blending of stains on the same tissue cross-section on demand. (b) A second field of view demonstrating the three digital stains generated using a single trained network. Contrast enhanced unstained tissue images are used for visual guidance; unprocessed raw versions of these images are used as input to the neural network. N/A (not available) refers to the fact that once a tissue section is histochemically stained with one type of stain, we cannot stain it with other stains subsequently; therefore, the comparison has N/A entries.*

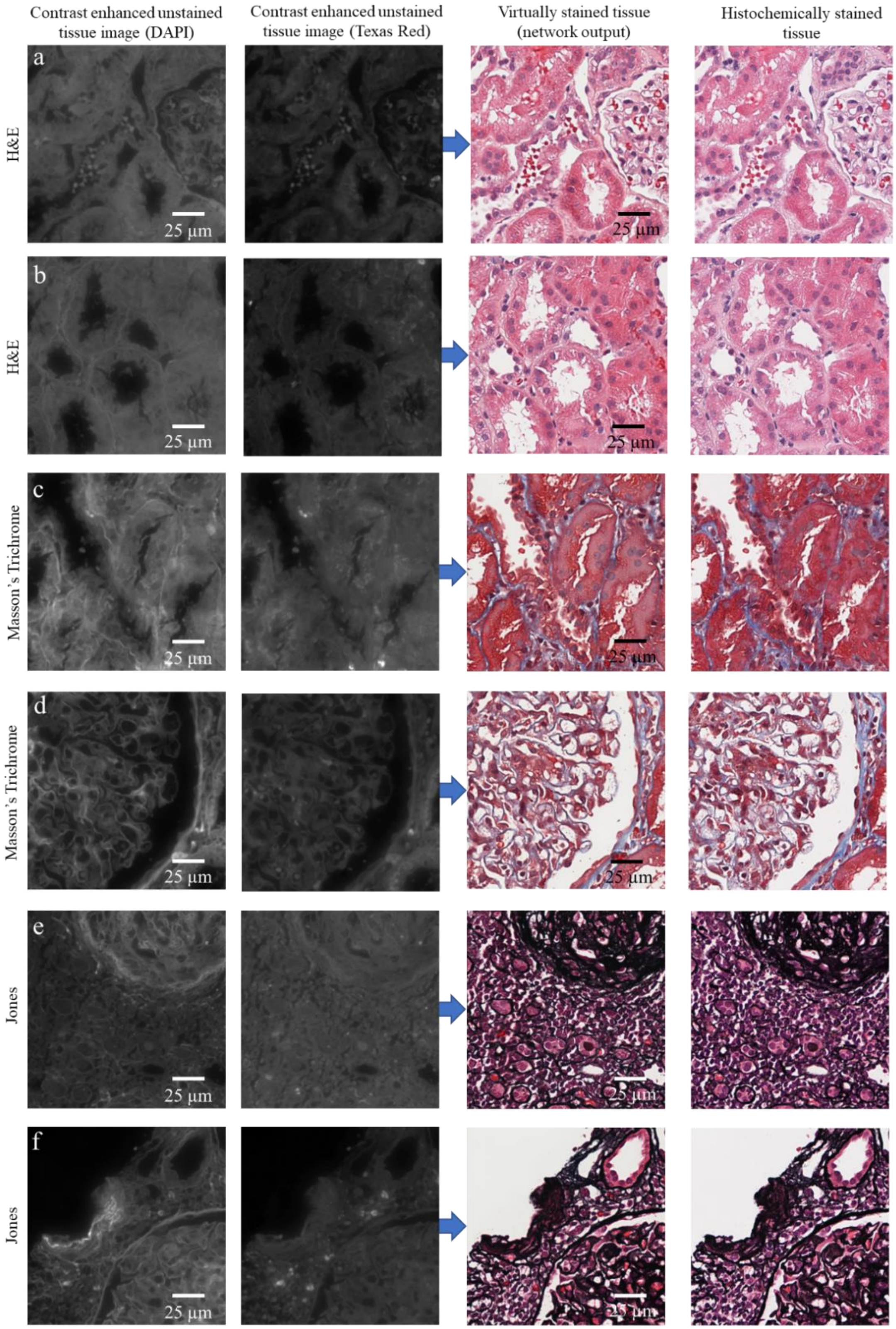

***Figure 2:*** *Examples of various fields of view which have been virtually stained using the presented multistain network. Co-registered histochemically stained fields of view of the same samples are also*

*shown to the right, and the unstained autofluorescence images are shown to the left in order to allow for direct comparison. (a,b) tissue stained with H&E, (c,d) tissue stained with Masson's trichrome, (e,f) tissue stained with Jones. Contrast enhanced unstained tissue images are used for visual guidance; unprocessed raw versions of these images are used as input to the neural network.*

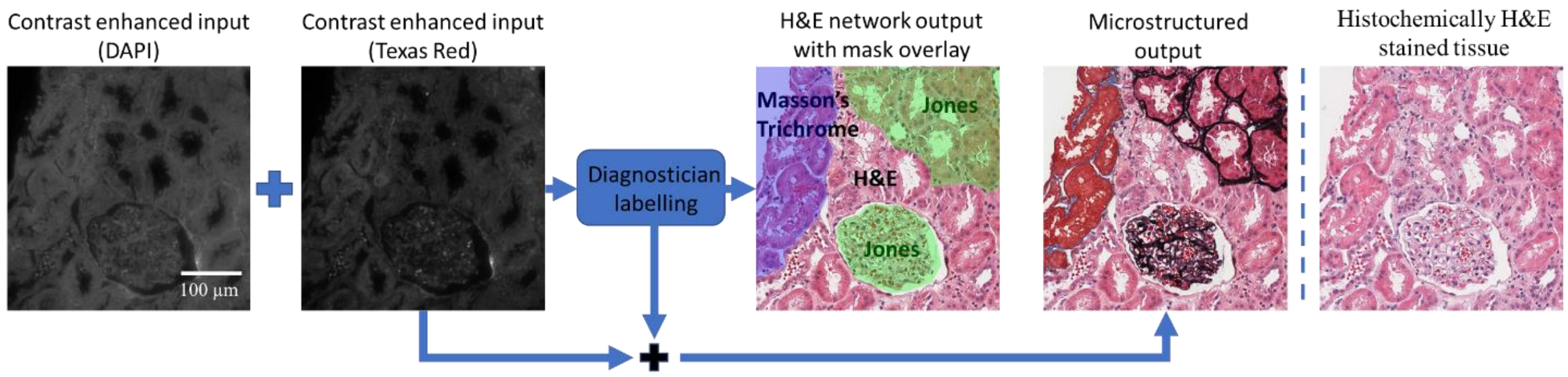

***Figure 3:*** *Example of multi-stain micro-structuring. A diagnostician or an algorithm can label sub-regions of the unstained tissue, creating on demand a digital staining matrix that defines the microscopic map of multiple stains to be virtually generated on the same tissue section. These labels are used by a single trained network to stain different areas of the tissue with the desired stains. A co-registered image of the histochemically stained H&E tissue (same sample) is shown for comparison. A histochemically stained image having the same or a similar microscopic map, with multiple stains on the same tissue section is not possible with today's chemical staining technology.*

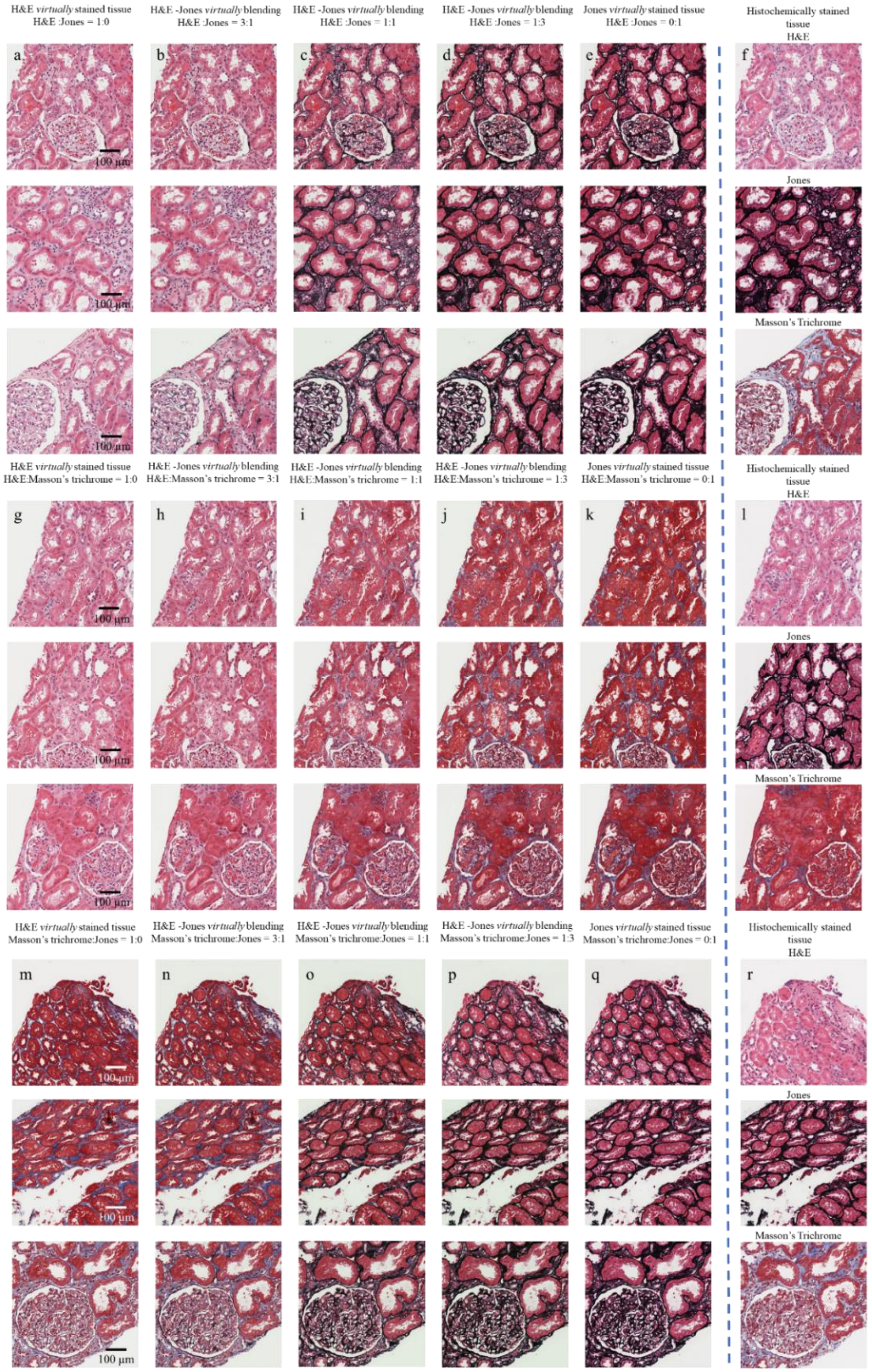

***Figure 4:*** *Examples of stain blending. (a-e) Kidney tissue that is virtually stained with varying class condition ratios of H&E to Jones stain. (g-k) Kidney tissue that is virtually stained with varying class*

*condition ratios of H&E to Masson's trichrome stain. (m-q) Kidney tissue that is virtually stained with varying class condition ratios of Masson's trichrome to Jones stain. (f,l,r) Co-registered images of the histochemically stained tissues (same samples) are shown for comparison (top H&E, middle Jones stain, bottom Masson's trichrome stain).*

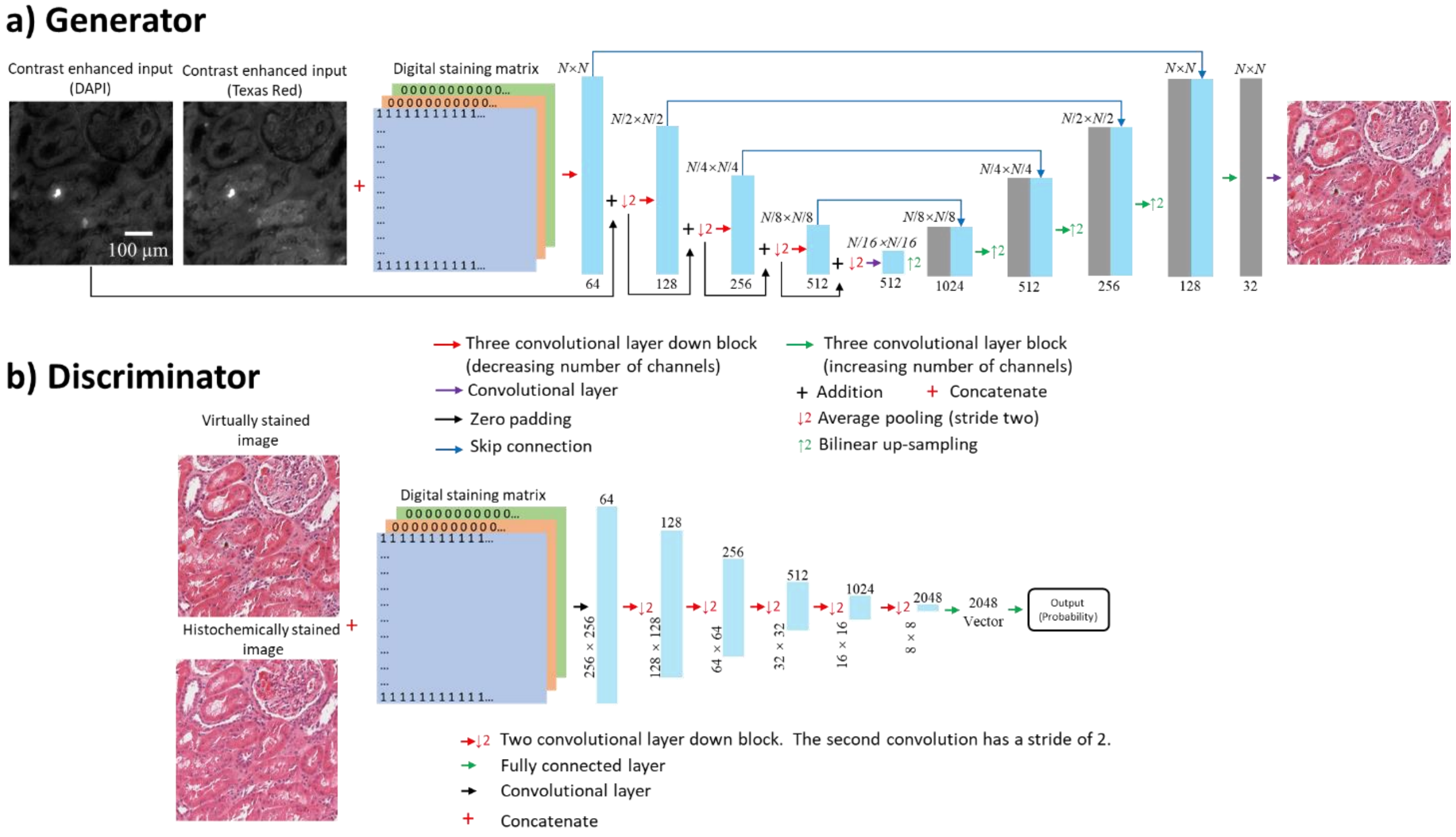

***Figure 5:*** *Diagram showing the network architecture of the GAN used to perform the transformation. (a) Generator network. (b) Discriminator network.*